\documentclass[journal,twoside]{IEEEtran}
\usepackage{textcomp}
\usepackage{stfloats}
\usepackage{url}
\usepackage{verbatim}
\usepackage{booktabs}
\usepackage{multirow}
\usepackage{xcolor}
\usepackage{cite}
\usepackage{latexsym}
\usepackage{graphicx}
\usepackage{subcaption}
\usepackage{amsfonts,amssymb,amsmath}
\usepackage{nccmath}
\usepackage{optidef}
\usepackage{hyperref}
\usepackage{color,soul}
\usepackage{array}
\usepackage{algorithm}
\usepackage{algorithmic}
\usepackage[T1]{fontenc}
\usepackage[utf8]{inputenc}
\usepackage{lastpage}
\usepackage{caption}
\usepackage{setspace}
\usepackage{bm}
\usepackage[nolist]{acronym}
\usepackage{mathtools}
\usepackage{tabularx}
\usepackage{steinmetz}
\usepackage{stackengine,scalerel}
\usepackage[makeroom]{cancel}
\usepackage{cuted}
\usepackage{afterpage}
\usepackage{lipsum,multicol}
\usepackage{tikz}	
\usetikzlibrary{arrows.meta}	
\usepackage{pgfplots} 
\usetikzlibrary{shadows,shadows.blur}	
\usetikzlibrary{plotmarks}
\usetikzlibrary{shapes.geometric}
\usepackage{siunitx}
\usepackage{titlesec}
\DeclareMathOperator{\trace}{tr}

\allowdisplaybreaks
\begin{document}
\title{Hybrid Radar Fusion with Quantization: CRB-Rate Trade-offs and ADC Dynamic Range}
\author{\IEEEauthorblockN{Akhileswar Chowdary$^{\dagger}$, Ahmad Bazzi$^{*\dagger}$, Marwa Chafii$^{*\dagger}$}\\
\IEEEauthorblockA{$^{\dagger}$NYU WIRELESS, NYU Tandon School of Engineering, Brooklyn, NY, USA, akhileswar.chowdary@nyu.edu\\$^{*}$Engineering division, New York University Abu Dhabi, UAE, \{ahmad.bazzi, marwa.chafii\}@nyu.edu}
}
\markboth{Accepted in IEEE Global Communications Conference (GLOBECOM) 2025}{Chowdary \MakeLowercase{\emph{et al.}}: Hybrid Radar Fusion with Quantization: CRB-Rate Trade-offs and ADC Dynamic Range}
\maketitle
\begin{abstract}
Recent advancements have underscored the relevance of low-resolution analog-to-digital converters (ADCs) in integrated sensing and communication (ISAC) systems. Nevertheless, their specific impact on hybrid radar fusion (HRF) remains largely unexplored. In HRF systems, where uplink (UL) paths carry direct and reflected signals in the same frequency band, the reflected signal is often significantly weaker, making HRF performance particularly sensitive to ADC resolution. To study this effect, we use the quantized Cramér-Rao bound (CRB) to measure sensing accuracy. This work derives an upper bound on the quantized CRB for angle of arrival (AoA) estimation and explores CRB-rate trade-offs through two formulated optimization problems. Simulation results indicate that HRF becomes infeasible when the dynamic range of the received signal exceeds the dynamic range supported by the ADC, which is inherently limited by its resolution. Furthermore, the UL communication rate does not increase significantly when the ADC resolution is raised beyond a certain threshold. These observations highlight a fundamental trade-off between sensing and communication performance: while HRF performance benefits from higher ADC resolutions, the corresponding gains in communication rate plateau. This trade-off is effectively characterized using CRB-rate boundaries derived through simulation.
\end{abstract}
\section{Introduction}\label{sec:intro}
\IEEEPARstart{A}{dvancements} in wireless communication, from the second generation (2G) to fifth generation (5G), have enabled transformative applications such as intelligent transportation \cite{zhang2025astars}, remote healthcare, UAV networks, and smart cities, which increasingly rely on integrated sensing and communication (ISAC) capabilities \cite{chafii_twelve_2023}. Recognizing ISAC’s critical role, the International Telecommunication Union (ITU) has designated ISAC as a key use case for 6G \cite{ITU_recommendation}, that led to a growing interest in dual-function radar-communication (DFRC) systems, which integrate radar and communication functionalities by sharing hardware \cite{10496165}, waveform \cite{bomfinmono24} and spectrum, resulting in reduced costs and improved spectral efficiency \cite{liu_seventy_2023}.

Despite these benefits, DFRC systems operating at millimeter-wave (mmWave) and sub-THz frequencies face significant challenges, especially with respect to hardware complexity and power consumption. High-resolution analog-to-digital converters (ADCs), required for these systems, contribute significantly to power consumption, which scales exponentially with ADC resolution and sampling rate \cite{761034}. Reducing ADC resolution has been proposed as an effective approach to mitigate these costs \cite{dutta_case_2020}.

Several studies have focused on low-resolution ADCs for ISAC systems. The authors in \cite{cheng_transmit_2021} have examined DFRC systems where one-bit digital-to-analog converters (DACs) are used at the transmitter and infinite-resolution ADCs are deployed at the receiver, optimizing communication performance under sensing constraints. Similarly, \cite{dizdar_rate-splitting_2021} has proposed rate-splitting multiple access (RSMA) with low-resolution DACs, balancing radar beamforming and DAC distortion. Other works have explored quantizer designs and hybrid analog/digital processing approaches for low-resolution systems, demonstrating significant performance improvements \cite{ruan_task-based_2024}. These solutions often aim to minimize the mean squared error (MSE) of communication or target estimation under finite-resolution constraints.

In addition, several studies have explored low-resolution ADCs in MIMO-DFRC systems. These include joint analog and digital processing schemes \cite{zhu_low-complexity_2021}, energy-efficient detection techniques \cite{xia_energy-efficient_2019}, and optimization of hybrid analog/digital receivers \cite{ma_bit_2021}. Recent research has also focused on transmit waveform design, and joint transmit precoding with reconfigurable intelligent surfaces (RIS), aiming to minimize MSE and interference while maintaining communication quality of service (QoS) \cite{wu_quantized_2024}. These works underscore the benefits of low-resolution ADCs in mmWave ISAC systems \cite{kumari_low-resolution_2020}.

However, the specific effects of ADC resolution on hybrid radar fusion (HRF) systems \cite{10417003}, where direct and reflected uplink (UL) signals operate in the same frequency band, remain underexplored. In HRF, the reflected signal often has significantly lower power than the direct path, requiring high-resolution ADCs to differentiate between them. Existing ISAC solutions are not directly applicable to HRF due to differences in system models and the lack of studies addressing the dynamic range (DR) needed for ADCs in HRF.

This paper investigates the impact of ADC resolution on HRF systems, particularly focusing on how the required resolution changes based on the power difference between direct and reflected signals. The study provides insights into selecting the appropriate ADC resolution for HRF systems under different power conditions, balancing performance with hardware efficiency and necessary DR. The main contributions of this paper are as follows.
\begin{itemize}
    \item \textbf{Upper Bound on CRB for AoA Estimation}: We derive an upper bound on the Cramér-Rao Bound (CRB) for angle of arrival (AoA) estimation in HRF systems, considering the effects of finite-resolution ADCs.
    \item \textbf{Characterization of CRB-Rate Trade-offs}: Using the derived upper bound on the CRB for AoA, we formulate optimization problems to characterize the CRB-rate boundary, allowing us to analyze the impact of ADC DR and quantization on AoA estimation in HRF systems.
    \item \textbf{Simulation Results}: Extensive simulations demonstrate how ADC DR and resolution influence AoA estimation performance in HRF systems, showcasing the effects of system parameters on both CRB and communication rates through the CRB-rate boundary.
\end{itemize}
This research examines the CRB-rate trade-off in HRF systems, emphasizing how ADC DR simultaneously affects sensing and communication performance. While most existing studies focus on downlink (DL) CRB-rate trade-offs, this work addresses the unique CRB-UL rate trade-off in HRF systems. To our knowledge, this is the first study to comprehensively analyze the effects of ADC DR and quantization on HRF systems. We hope this research serves as a foundation for future work in this area, particularly in the presence of hardware impairments, and aids in developing strategies to mitigate such challenges.
\section{System Model}\label{sec:sys_mod}
\begin{figure}[!t]
    \centering
    \includegraphics[width=0.8\columnwidth]{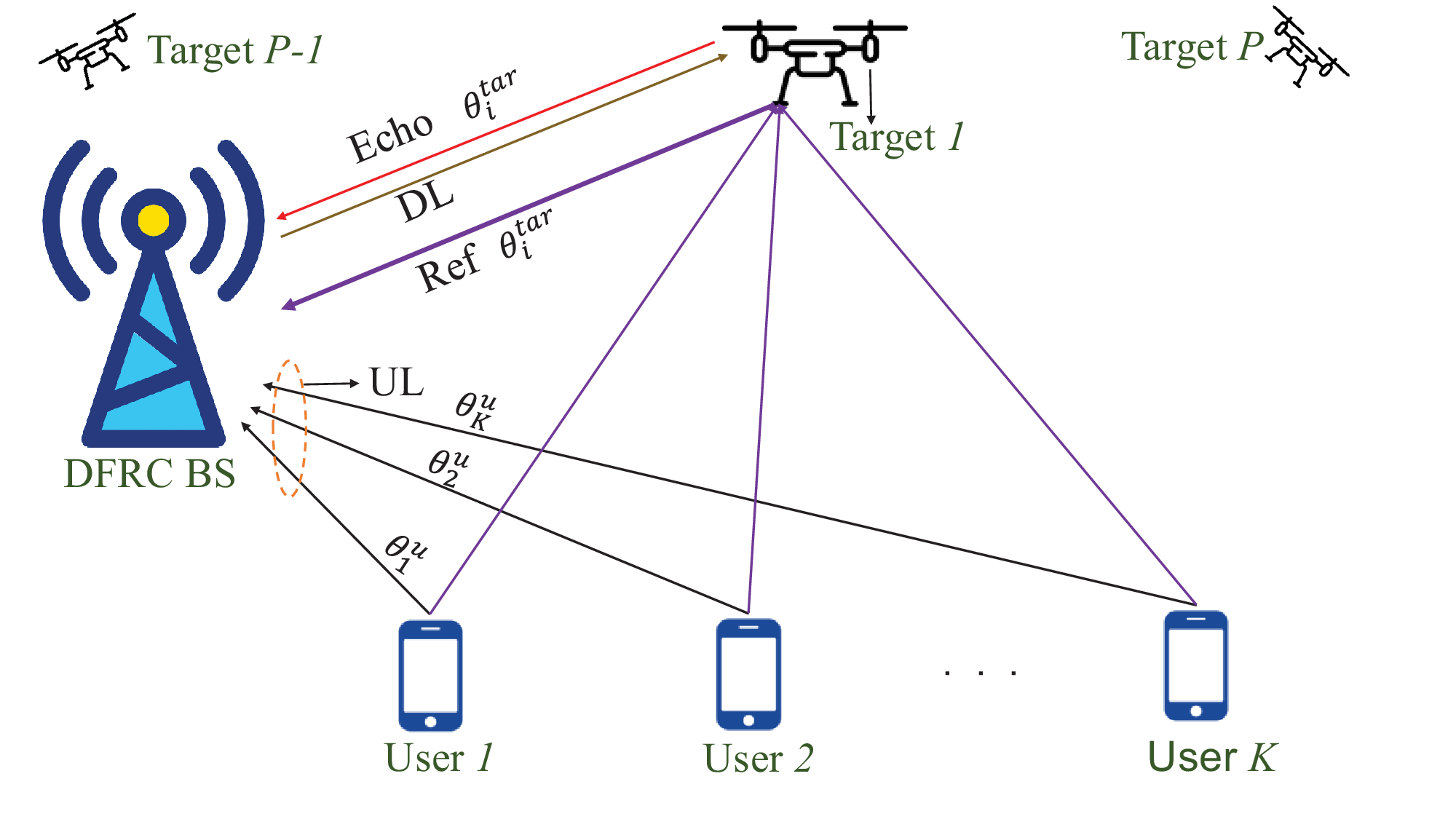}\vspace{-0.1in}
    \caption{HRF system consisting of a DFRC BS operating in monostatic mode, receiving echoes from $P$ radar targets, and uplink signals from $K$ users.}
    \label{fig:system_model}
\end{figure}

Fig.~\ref{fig:system_model} depicts the configuration of the HRF system, where a DFRC BS serves $K$ communication users, represented by the set $\mathcal{K} \in \{1,\ldots,K\}$, with $P$ targets in the environment. The DFRC BS is equipped with $N$ transmit and receive antennas. Target $i$ is located at an angle $\theta_{i}^{\mathrm{tar}}$ relative to the DFRC BS, with the collective AoAs denoted as $\Theta^{\mathrm{tar}} = \{\theta_{1}^{\mathrm{tar}},\ldots,\theta_{P}^{\mathrm{tar}}\}$. In the DL, the $\ell^{\mathrm{th}}$ transmitted OFDM symbol is expressed as \useshortskip
\begin{equation}
    \mathbf{s}_{\ell,0}(t) = \left(\sum_{m \in \mathcal{C}_0} b_{m,0}^{(\ell)} e^{j 2 \pi m \Delta_f t} \Pi(t - \ell T) \right) \mathbf{f},
\end{equation}
where $\mathbf{s}_{\ell,0}(t) \in \mathbb{C}^{N\times1}$ represents the transmitted signal vector, $\mathcal{C}_{0}$ is the subcarrier set for DL, $b_{m,0}^{(\ell)}$ modulates the $m^{\mathrm{th}}$ subcarrier, $\Delta_f$ is the subcarrier spacing, $\Pi(t)$ is a windowing function, $T = 1/\Delta_f$ is the symbol duration, and $\mathbf{f} \in \mathbb{C}^{N \times 1}$ is the precoding vector. Similarly, in the UL, the $\ell^{\mathrm{th}}$ OFDM symbol transmitted by user $k$ with $N_{k}^{\mathrm{u}}$ antennas is given by \useshortskip
\begin{equation}
     \mathbf{s}_{\ell,k}(t) = \left(\sum_{m \in \mathcal{C}_k} b_{m,k}^{(\ell)} e^{j 2 \pi m \Delta_f t} \Pi(t - \ell T) \right) \mathbf{f}_{k}^\mathrm{u},
\end{equation}
where $\mathbf{s}_{\ell,k}(t) \in \mathbb{C}^{N_{k}^{\mathrm{u}} \times 1}$, $\mathcal{C}_{k}$ is the subcarrier set assigned to user $k$, $b_{m,k}^{(\ell)}$ is the modulating symbol for subcarrier $m$, and $\mathbf{f}_{k}^{\mathrm{u}} \in \mathbb{C}^{N_{k}^{\mathrm{u}} \times 1}$ is the user's precoding vector. These signals are then upconverted to the center frequency $f_{\mathrm{c}}$ before transmission. The DFRC BS receives and downconverts the signal, represented by $\mathbf{r}_{\ell}(t) = \mathbf{r}_{\ell}^{\mathrm{echo}}(t) + \sum_{k=1}^{K} \mathbf{r}_{\ell,k}^{\mathrm{UL}}(t) + \mathbf{z}_{\ell}(t)$, where $\mathbf{r}_{\ell}^{\mathrm{echo}}(t)$ is the received radar echo, $\mathbf{r}_{\ell,k}^{\mathrm{UL}}(t)$ is the signal from user $k$, and $\mathbf{z}_{\ell}(t)$ denotes AWGN with zero mean and variance $\sigma^2$. The expressions for $\mathbf{r}_{\ell}^{\mathrm{echo}}$ and $\mathbf{r}_{\ell,k}^{\mathrm{UL}}$ are as follows \useshortskip
\begin{flalign}
    &\mathbf{r}_{\ell}^{\mathrm{echo}}(t) = \sum_{i \in \Phi_{0}} \gamma_{i,\ell}^{\mathrm{tar}} \mathbf{a}_{r}(\theta_{i}^{\mathrm{tar}}) \mathbf{a}_{t}^{\mathrm{T}}(\theta_{i}^{\mathrm{tar}}) \mathbf{s}_{\ell,0}(t - 2 \tau_{i}^{\mathrm{tar}}),&& \label{eq:echo-t} \\
    &\mathbf{r}_{\ell,k}^{\mathrm{UL}}(t) = \gamma_{k,0}^{\mathrm{u}} \mathbf{a}_{r}(\theta_{r,k}^{\mathrm{u}})(\mathbf{a}_{k,t}^{\mathrm{u}}(\theta_{k}^{\mathrm{u}}))^{\mathrm{T}}\mathbf{s}_{k,\ell}(t - \tau_{k}^{\mathrm{u}})&& \label{eq:UL-t} \\
    &+ \sum_{j \in \Phi_{k}} \gamma_{k,j,\ell}^{\mathrm{tar}} \mathbf{a}_{r}(\theta_{j}^{\mathrm{tar}})(\mathbf{a}_{k,t}^{\mathrm{u}}(\theta_{k,j}^{\mathrm{u}}))^{\mathrm{T}} \mathbf{s}_{k,\ell}(t - \phi_{k,j}).&& \nonumber
\end{flalign}
In these equations, $\Phi_0$ and $\Phi_k$ are the target sets observed by the DFRC BS and user $k$, respectively, with $\Phi_k = [u_1^{(k)}, \ldots , u^{(k)}_{\vert \Phi_k \vert}]$ for each $k \in \mathcal{K}$. The union of these sets encompasses all $P$ targets, and each user generally observes fewer targets than the DFRC BS, i.e., $\vert \Phi_k \vert \leq \vert \Phi_0 \vert$ with non-overlapping target sets for different users. Further, $\gamma_{i,\ell}^{\mathrm{tar}} = g_{i}^{\mathrm{tar}} e^{-j2\pi f_{c} 2\tau_{i}^{\mathrm{tar}}} e^{j2\pi f_{D,i} \ell T}$, where $g_{i}^{\mathrm{tar}}$ is the channel gain for target $i$, $\tau_{i}^{\mathrm{tar}}$ is the round-trip delay, and $f_{D,i}$ represents Doppler shift. Similarly, $\gamma_{k,0}^{\mathrm{u}} = g_{k,0}^{\mathrm{u}} e^{-j2\pi f_{c} \tau_{k}^{\mathrm{u}}}$, with $g_{k,0}^{\mathrm{u}}$ as the gain between user $k$ and the DFRC BS, $\tau_{k}^{\mathrm{u}}$ as the delay, and $\theta_{k}^{\mathrm{u}}$ and $\theta_{r,k}^{\mathrm{u}}$ as the angles of departure (AoD) and AoA. Other parameters follow similarly, with $\mathbf{a}_{r}(\theta)$ as the DFRC BS receiving steering vector, $\mathbf{a}_{t}(\theta)$ as the BS transmit steering vector, and $\mathbf{a}_{k,t}^{\mathrm{u}}(\theta)$ as the user’s transmit steering vector. Using a \textit{narrowband assumption}, the response for the $n^{\mathrm{th}}$ antenna at angle $\theta$ is $a_n(\theta) = \exp(j2\pi \frac{d}{\lambda} (n-1) \sin(\theta))$.

A $b-$bit ADC then processes the received signal with pre-filtering, sampling at intervals $v \triangleq v \frac{T}{M}$, and quantization. The sampled signal can be written as $\mathbf{r}_{\ell}[v] = \mathbf{x}_{\ell}[v] + \mathbf{z}_{\ell}[v]$, where \useshortskip
\begin{align}\label{eq:x-ell-v}
    \mathbf{x}_{\ell}[v] &= \sum\limits_{\substack{i\in\Phi_{0},\\ m_1\in\mathcal{C}_0}} b^{\ell}_{m_1,0}\mathbf{H}_{\ell,m_1,i}^{\mathrm{echo}}[v]\mathbf{f} + \sum\limits_{\substack{k=1,\\ m_2\in\mathcal{C}_{k}}}^{K}b^{\ell}_{m_2,k}\mathbf{H}_{\ell,m_2,k}^{\mathrm{dp}}[v]\mathbf{f}_{k}^{\mathrm{u}}\nonumber\\ 
    &+\sum\nolimits_{\substack{k=1,\\ m_2\in\mathcal{C}_{k}}}^{K}\sum\nolimits_{j\in\Phi_{k}} b^{\ell}_{m_2,k}\mathbf{H}_{\ell,m_2,k,j}^{\mathrm{ref}}[v]\mathbf{f}_{k}^{\mathrm{u}}.\nonumber
\end{align}
Here, $\mathbf{r}_{\ell}[v] \in \mathbb{C}^{N \times 1}$ is the DFRC BS’s received signal vector, with $\mathbf{H}_{\ell,m_1,i}^{\mathrm{echo}}$, $\mathbf{H}_{\ell,m_2,k}^{\mathrm{dp}}$, and $\mathbf{H}_{\ell,m_2,k,j}^{\mathrm{ref}}$ representing backscattered, direct path, and reflected channels, respectively. Their expressions are \useshortskip
\begin{align}
    &\mathbf{H}_{\ell,m,i}^{\mathrm{echo}}[v] = \Tilde{\gamma}_{i,\ell}^{\mathrm{tar}} c_{m}(2\tau_{i}^{\mathrm{tar}},v) \mathbf{a}_{r}(\theta_{i}^{\mathrm{tar}}) (\mathbf{a}_{t}(\theta_{i}^{\mathrm{tar}}))^{\mathrm{T}},\\
    &\mathbf{H}_{\ell,m,k}^{\mathrm{dp}}[v] = g_{k,0}^{\mathrm{u}} c_{m}(\tau_{k}^{\mathrm{u}},v) \mathbf{a}_{r}(\theta_{r,k}^{\mathrm{u}}) (\mathbf{a}_{k,t}^{\mathrm{u}}(\theta_{k}^{\mathrm{u}}))^{\mathrm{T}},\\
    &\mathbf{H}_{\ell,m,k,j}^{\mathrm{ref}}[v] = \Tilde{\gamma}_{k,j,\ell}^{\mathrm{tar}} c_{m}(\phi_{k,j},v) \mathbf{a}_{r}(\theta_{j}^{\mathrm{tar}}) (\mathbf{a}_{k,t}^{\mathrm{u}}(\theta_{k,j}^{\mathrm{u}}))^{\mathrm{T}},
\end{align}
where $\Tilde{\gamma}_{i,\ell}^{\mathrm{tar}} = g_{i}^{\mathrm{tar}}e^{j2\pi f_{D,i}\ell T}$, $c_m(\tau,v) = e^{-j2\pi ((m\Delta_f+f_{c}) \tau - mv/M)}$, $\Tilde{\gamma}_{k,j,\ell}^{\mathrm{tar}} = g_{k,j}^{\mathrm{tar}}e^{j2\pi f_{D,i}\ell T}$. Additionally, the UL channel $\mathbf{H}_{\ell,m,k}^{\mathrm{UL}}$ for user $k$ at the DFRC BS is represented as $\mathbf{H}_{\ell,m,k}^{\mathrm{UL}} = \mathbf{H}_{\ell,m,k}^{\mathrm{dp}} + \sum_{j=1}^{P} \mathbf{H}_{\ell,m,k,j}^{\mathrm{ref}}$. The sampled signal $\mathbf{r}_{\ell}[v]$ is quantized through a $b$-bit scalar quantizer, where real and imaginary parts are quantized separately, producing the output $\mathbf{r}_{\ell}^{\mathrm{q}}[v] = \mathcal{Q}(\mathbf{r}_{\ell}[v])$, assuming Gaussian-distributed inputs and optimal quantization.
\section{Quantized CRB and lower bound on FIM}\label{sec:crb_der}
In this section, we present the expressions of the quantized CRB for HRF and a lower bound on the FIM. These expressions form the foundation for analyzing the performance of quantized HRF systems, particularly regarding the trade-offs between the CRB and the communication rate. We consider the quantized signal $\mathbf{r}_{\ell}^{\mathrm{q}}$, where the goal is to estimate one or more than one unknown deterministic parameters of interest in the vector, $\pmb{\psi} = \left[\theta^{\mathrm{tar}}_{i}, f_{D,i}, \tau^{\mathrm{tar}}_{i}, \theta_{r,k}^{\mathrm{u}}, \theta^{\mathrm{u}}_{k}, \tau^{\mathrm{u}}_{k}, \phi_{k,j}, \theta_{k,j}^{\mathrm{u}}, g_{i}^{\mathrm{tar}}, g_{k,0}^{\mathrm{u}}, g_{k,j}^{\mathrm{u}}\right]$ for each $k \in \mathcal{K}$ and $i, j \in \Phi_{0}$. To this end, we compute the likelihood function for the quantized signal, represented as \( p\left(\mathbf{r}^{\mathrm{q}}_{\ell}|\mathbf{x}_{\ell}(\pmb{\psi})\right) \) and is expressed as follows
\begin{flalign}
    &p\left(\mathbf{r}^{\mathrm{q}}_{\ell}|\mathbf{x}_{\ell}(\pmb{\psi})\right) = \prod\nolimits_{n=1}^{N}p\left(r^{\mathrm{q}}_{n,\ell}|x_{n,\ell}(\pmb{\psi})\right) \nonumber &&\\
    &= \prod\nolimits_{n=1}^{N}p\left(u_{n,\ell}|x_{n,\ell}(\pmb{\psi})\right)\prod\nolimits_{n=1}^{N}p\left(w_{n,\ell}|x_{n,\ell}(\pmb{\psi})\right),&&
\end{flalign}
where $r^{\mathrm{q}}_{n,\ell}$, $u_{n,\ell}$, and $w_{n,\ell}$ are the $n^{\mathrm{th}}$ elements of $\mathbf{r}_{\ell}^{\mathrm{q}}$, $\mathcal{Q}(\Re[\mathbf{r}_{\ell}])$, and $\mathcal{Q}(\Im[\mathbf{r}_{\ell}])$, respectively. We refer readers to \cite{chowdary2024fundamentaltradeoffsquantizedhybrid} for a complete derivation of the FIM for each parameter in $\pmb{\psi}$. Here, we provide the final FIM expression using the FIMs of real and imaginary parts of the quantized signal. The expressions are given as follows
\begin{flalign}
    &\left[\mathbf{F}_{\mathbf{u}}\right]_{ij} = \sum_{\ell=1}^{L} \sum_{n=1}^{N} \sum_{b=1}^{B} \frac{2}{\sigma^2} \frac{\partial x_{n,\ell}^{\mathrm{R}}}{\partial \psi_{i}} \frac{\partial x_{n,\ell}^{\mathrm{R}}}{\partial \psi_{j}} \Lambda_{n,\ell,b}^{\mathrm{R}},&&\\
    &\left[\mathbf{F}_{\mathbf{w}}\right]_{ij} = \sum_{\ell=1}^{L} \sum_{n=1}^{N} \sum_{b=1}^{B} \frac{2}{\sigma^2} \frac{\partial x_{n,\ell}^{\mathrm{I}}}{\partial \psi_{i}} \frac{\partial x_{n,\ell}^{\mathrm{I}}}{\partial \psi_{j}} \Lambda_{n,\ell,b}^{\mathrm{I}},&&
\end{flalign}
where $L$ is the total number of OFDM symbols transmitted, $B = 2^b$, $\mathbf{u} = [u_{1,1},\ldots,u_{N,L}]^{\mathrm{T}}$, $\mathbf{w} = [w_{1,1},\ldots,w_{N,L}]^{\mathrm{T}}$, and $\Lambda_{n,\ell,b}^{\mathrm{y}} = \frac{\left[\varphi(\alpha_{n,\ell,b}^{\mathrm{y}}) - \varphi(\beta_{n,\ell,b}^{\mathrm{y}})\right]^2}{\Phi(\alpha_{n,\ell,b}^{\mathrm{y}})-\Phi(\beta_{n,\ell,b}^{\mathrm{y}})}$, $y \in \{\mathrm{R}, \mathrm{I}\}$, $\Phi(x) = \int\nolimits_{-\infty}^{x}\varphi(t)dt$, and $\varphi(x) = \left(1/\sqrt{2\pi}\right)e^{-x^2/2}$. By leveraging the additive property of the FIMs for the real and imaginary components, we combine them to obtain the overall FIM for the quantized signal, $\mathbf{r}^{\mathrm{q}}(\pmb{\psi}) = [(\mathbf{r}^{\mathrm{q}}_{1})^{\mathrm{T}},\ldots,(\mathbf{r}^{\mathrm{q}}_{L})^{\mathrm{T}}]^{\mathrm{T}}$ as
\begin{equation}\label{eq:finalbbitFIM}
    \left[\mathbf{F}_{\mathbf{r}^{\mathrm{q}}}\right]_{ij} = \left[\mathbf{F}_{\mathbf{u}}\right]_{ij} + \left[\mathbf{F}_{\mathbf{w}}\right]_{ij}.
\end{equation}
The CRB for the parameter $\psi_{i}$, denoted as $\text{CRB}_{\psi_{i}}$, is obtained as the inverse of the FIM, \(\text{CRB}_{\psi_{i}} = \left(\left[\mathbf{F}_{\mathbf{r}^{\mathrm{q}}}(\pmb{\psi})\right]^{-1}\right)_{ii}\). To analyze the impact of the ADC DR on HRF performance, particularly through CRB-rate boundaries, we formulate optimization problems. However, using the exact FIM in \eqref{eq:finalbbitFIM} for optimization is highly challenging due to its complexity. Given the dual objective of optimizing the sensing CRB and communication rates, deriving lower bounds on the FIM and the rate offers a more tractable approach. To this end, we use the Bussgang theorem \cite{jjbussgang_1952} to represent the quantized signal as a linear model, where $\mathbf{r}^{\mathrm{q}}_{\ell} = \mathbf{G}\mathbf{r}_{\ell} + \mathbf{z}^{\mathrm{q}}_{\ell} = \mathbf{G}\mathbf{x}_{\ell} + \mathbf{G}\mathbf{z}_{\ell} + \mathbf{z}^{\mathrm{q}}_{\ell} = \mathbf{G}\mathbf{x}_{\ell} + \Bar{\mathbf{z}}_{\ell}$. Here, $\mathbf{G}$ is the distortion matrix, $\mathbf{z}^{\mathrm{q}}_{\ell}$ represents the uncorrelated quantization noise, and $\Bar{\mathbf{z}}_{\ell}$ is the effective noise after quantization. Detailed derivations of $\mathbf{G}$, the covariance matrix of the quantization noise $\Bar{\mathbf{z}}_{\ell}$, and the resulting lower bound on the FIM are presented in \cite{chowdary2024fundamentaltradeoffsquantizedhybrid}. The final expression of the lower bound on FIM is given as\useshortskip
\begin{equation}
\scalebox{1}{$
\left[\mathbf{F}_{\mathbf{r}^{\mathrm{q}}}^{\text{LS}}\right]_{ij} \geq 2\Re\left[\sum_{\ell = 1}^{L} \left(\frac{\partial\mathbf{G}\mathbf{x}_{\ell}}{\partial\psi_{i}}\right)^{\mathrm{H}} \left(\mathbf{R}_{\Bar{\mathbf{z}}_{\ell}\Bar{\mathbf{z}}_{\ell}}^{\mathrm{LS}}\right)^{-1} \left(\frac{\partial\mathbf{G}\mathbf{x}_{\ell}}{\partial\psi_{j}}\right)\right],
$}
\end{equation}
where $\mathbf{R}_{\Bar{\mathbf{z}}_{\ell}\Bar{\mathbf{z}}_{\ell}}^{\mathrm{LS}}$ is the covariance matrix of the effective noise with low per-antenna SNR approximation, $\mathrm{LS}$ denotes the low SNR assumption. When $\mathbf{R}_{\mathbf{z}_{\ell}\mathbf{z}_{\ell}} = \sigma^2 \mathbf{I}_{N}$, the covariance simplifies to $\mathbf{R}_{\Bar{\mathbf{z}}_{\ell}\Bar{\mathbf{z}}_{\ell}}^{\mathrm{LS}} = \sigma^2(1 - \eta)\mathbf{I}_{N}$, leading to the simplified bound\useshortskip
\begin{flalign}\label{eq:lowsnrfimbound1}
    &\scalebox{0.895}{$\left[\mathbf{F}_{\mathbf{r}^{\mathrm{q}}}^{\text{LS}}\right]_{ij} \geq \frac{2(1 - \eta)}{\sigma^2}\Re\left[\sum_{\ell = 1}^{L} \sum_{n=1}^{N} \left(\frac{\partial x_{n,\ell}(\pmb{\psi})}{\partial\psi_{i}}\right)^{*} \left(\frac{\partial x_{n,\ell}(\pmb{\psi})}{\partial\psi_{j}}\right)\right].$}&&
\end{flalign}

% \section{Lower Bound on FIM of HRF}\label{sec:lb_FIM}
% \input{Sections/lower_bound_FIM}
\section{Lower Bound on Rate and 1-bit Quantization}\label{sec:lb_rate}
In this section, we present the expressions of the lower bound on the UL rate. We use the result from \cite{diggavi_worst_2001} that mutual information (MI) is minimized under Gaussian-distributed effective noise to establish the lower bound on the UL rate. Using the approximate covariance matrices derived earlier, the lower bound on the UL rate is given as follows \cite{chowdary2024fundamentaltradeoffsquantizedhybrid}.
\begin{equation}\label{eq:mutualinflb1}
    I(\mathbf{s}_{\ell};\mathbf{r}^{\mathrm{q}}_{\ell}) \geq \log_{2}\left|\mathbf{I}_{N} + \left(\frac{1-\eta}{\sigma^2}\right)\mathbf{R}_{\mathbf{x}_{\ell}\mathbf{x}_{\ell}}\right|.
\end{equation}
However, when we consider 1-bit quantization, we have exact expressions for the covariance matrix of the effective noise, which is derived through the arcsine law. The covariance matrices of the quantized signal, effective noise after quantization in the case of 1-bit quantizers, are given in \cite{chowdary2024fundamentaltradeoffsquantizedhybrid}. However, the final expression of the MI for 1-bit ADCs is given as
\begin{equation}\label{eq:mutualinf1bitlowsnr}
I(\mathbf{s}_{\ell};\mathbf{r}^{\mathrm{q}}_{\ell})^{\mathrm{1-bit}} \approx \frac{2}{\pi\sigma^2} \text{tr}\left(\mathbf{R}_{\mathbf{x}_{\ell}\mathbf{x}_{\ell}}\right).
\end{equation}
% \subsection{1-bit Quantization}\label{sec:1bit}
% \input{Sections/1bit_quantization}
\section{Optimization framework for CRB-UL Rate trade-off characterization}\label{sec:optmization}
In this section, we formulate two optimization problems, labeled $\mathbb{P}_{0}$ and $\mathbb{P}_{1}$, to systematically analyze the CRB-rate trade-off boundary in HRF systems. This boundary highlights the inherent trade-offs between CRB and communication rate in the HRF systems. To accurately represent this boundary, we employ optimization methods using the derived lower bounds on FIM and rate from \eqref{eq:lowsnrfimbound1} and \eqref{eq:mutualinflb1}, respectively, for computing the CRB and rate. These optimization problems are essential, as each endpoint on the CRB-rate boundary corresponds to minimizing the CRB or maximizing the rate. For complete boundary characterization, however, we balance these objectives by imposing constraints on one parameter while optimizing the other. To this end, we define two optimization problems: $\mathbb{P}_{0}$ minimizes the CRB for a specific sensing parameter while ensuring that the communication rate meets a required minimum quality of service (QoS) for the user, whereas $\mathbb{P}_{1}$ aims to maximize the communication rate while maintaining a specified sensing accuracy.

To study the effects of ADC DR on AoA estimation, we focus on formulating the optimization problem specifically for the AoA of the target, denoted as $\theta_{i}^{\mathrm{tar}}$. Similar optimization problems can also be formulated for other parameters of interest.

\subsection{Sensing-Centric Design}\label{sec:sensing_centric}
The optimization problem $\mathbb{P}_{0}$ is structured as follows
\begin{subequations}
    \begin{eqnarray}
        (\mathbb P_0) \quad & \underset{\mathbf{f}_0, \{\mathbf{f}_k^{\mathrm{u}}\}_{k = 1}^K}{\operatorname{minimize}} &\left(\text{CRB}_{\theta_{i}^{\mathrm{tar}}}^{\mathrm{LS}}\left(\mathbf{f},\mathbf{f}_{k}^{\mathrm{u}}\right)\right) \label{eq:P0} \\
        &\operatorname{s.t.} & \trace\left(\mathbf{R}_{k}\right)\leq P^{\mathrm{u}}_{\mathrm{max}},\;\forall k \in \mathcal{K}\label{eq:P01}\\
        && \trace\left(\mathbf{R}_{0}\right)\leq P^{\mathrm{BS}}_{\mathrm{max}}\label{eq:P02}\\
        && \mathrm{rank}\left(\mathbf{R}_{0}\right) = 1\label{eq:P03}\\
        && \mathrm{rank}\left(\mathbf{R}_{k}\right) = 1,\;\forall k \in \mathcal{K}\label{eq:P04}\\
        &&I(\mathbf{s}_{\ell};\mathbf{r}^{\mathrm{q}}_{\ell})\geq \mu \label{eq:P05}
    \end{eqnarray}
\end{subequations}
where $\text{CRB}_{\theta_{i}^{\mathrm{tar}}}^{\mathrm{LS}} = \left[\left(\mathbf{F}_{\mathbf{r}^{\mathrm{q}}}^{\text{LS}}(\Theta)\right)^{-1}\right]_{ii}$ as in \eqref{eq:fimthetalowsnr}, $\mathbf{R}_{0} = \mathbf{f}\mathbf{f}^{\mathrm{H}}$, and $\mathbf{R}_{k} = \mathbf{f}_{k}^{\mathrm{u}}\left(\mathbf{f}_{k}^{\mathrm{u}}\right)^{\mathrm{H}}$, with $\mathbf{R}_{0}$ and $\mathbf{R}_{k}$ as positive semidefinite matrices. The objective in \eqref{eq:P0} is to minimize the CRB for the HRF system. Constraints \eqref{eq:P01} and \eqref{eq:P02} limit the power of the users and BS to $P^{\mathrm{u}}_{\mathrm{max}}$ and $P^{\mathrm{BS}}_{\mathrm{max}}$, respectively, representing quadratic constraints. The constraints in \eqref{eq:P03} and \eqref{eq:P04} enforce rank-1 conditions on the BS and user precoders, while the MI constraint \eqref{eq:P05} ensures a minimum communication rate. Constraints \eqref{eq:P03} and \eqref{eq:P04} are non-convex. To address this, we use rank-1 relaxation and reformulate $\mathbb{P}_{0}$ as follows.
\begin{subequations}
    \begin{eqnarray}
        (\mathbb{P}_0) \quad & \underset{\mathbf{R}_{0},\{\mathbf{R}_{k}\}_{k=1}^{K}}{\operatorname{minimize}} 
        & \left(\text{CRB}_{\theta_{i}^{\mathrm{tar}}}^{\mathrm{LS}}\left(\mathbf{R}_{0},\mathbf{R}_{k}\right)\right) \label{eq:P0_1} \\
        &\operatorname{s.t.} & \eqref{eq:P01}, \eqref{eq:P02}, \eqref{eq:P05},  \notag\\
        && \mathbf{R}_{0}\succeq 0\label{eq:P03_3}\\
        && \mathbf{R}_{k} \succeq 0,\;\forall k \in \mathcal{K}.\label{eq:P04_4}
    \end{eqnarray}
\end{subequations}
This reformulated convex problem can be solved using CVX, enabling us to find the minimum CRB under the specified power and rate constraints.

\subsection{Communication-Centric Design}\label{sec:comm_centric}
We next define the optimization problem $\mathbb{P}_{1}$, which maximizes the communication rate of the HRF system under a constraint on the CRB. The problem $\mathbb{P}_{1}$ is structured as follows.
\begin{subequations}
    \begin{eqnarray}
        (\mathbb{P}_1) \quad & \underset{\mathbf f_0, \{\mathbf f_k^{\mathrm{u}}\}_{k = 1}^K}{\operatorname{maximize}} & I(\mathbf{s}_{\ell};\mathbf{r}^{\mathrm{q}}_{\ell}) \label{eq:P1} \\
        &\operatorname{s.t.} & \left(\text{CRB}_{\theta_{i}^{\mathrm{tar}}}^{\mathrm{LS}}\left(\mathbf{f},\mathbf{f}_{k}^{\mathrm{u}}\right)\right)\leq \Gamma\label{eq:P11}\\
        && \eqref{eq:P01},\eqref{eq:P02},\eqref{eq:P03},\eqref{eq:P04}, \notag
    \end{eqnarray}
\end{subequations}
where $\Gamma$ denotes the CRB threshold. The expression for $I(\mathbf{s}_{\ell};\mathbf{r}^{\mathrm{q}}_{\ell})$ is provided in \eqref{eq:mutualinflb1}. As with $\mathbb{P}_{0}$, the non-convex constraints \eqref{eq:P03} and \eqref{eq:P04} are addressed using rank-1 relaxation, yielding the following reformulation.
\begin{subequations}
    \begin{eqnarray}
        (\mathbb{P}_1) \quad & \underset{\mathbf{R}_{0},\{\mathbf{R}_{k}\}_{k=1}^{K}}{\operatorname{minimize}} & I(\mathbf{s}_{\ell};\mathbf{r}^{\mathrm{q}}_{\ell}) \label{eq:P1_1} \\
        &\operatorname{s.t.} & \text{CRB}_{\theta_{i}^{\mathrm{tar}}}^{\mathrm{LS}}\left(\mathbf{R}_{0},\mathbf{R}_{k}\right)\leq \Gamma \label{eq:P11_1}\\
        && \eqref{eq:P01},\eqref{eq:P02},\eqref{eq:P03_3},\eqref{eq:P04_4}. \notag
    \end{eqnarray}
\end{subequations}
To determine the extreme values on the CRB-rate boundary, we set $\mu = 0$ in $\mathbb{P}_{0}$ and $\Gamma = 0$ in $\mathbb{P}_{1}$, solving these formulations with CVX. These two extreme points on the boundary represent the “sensing-centric” and “communication-centric” configurations, respectively. Specifically, with $\mu = 0$ in $\mathbb{P}_{0}$, we achieve a solution that minimizes the CRB upper bound, prioritizing sensing accuracy while allowing the lowest feasible rate. Conversely, setting $\Gamma = 0$ in $\mathbb{P}_{1}$ results in a solution that maximizes the communication rate while allowing for the maximum CRB. This setting prioritizes communication performance over sensing precision. We generate the CRB-rate boundary by varying the parameter $\mu$ in $\mathbb{P}_{1}$ between these extreme points, thus mapping the trade-offs between sensing accuracy and communication throughput. This method provides an in-depth understanding of the performance limits of the HRF system under different ADC resolutions and power constraints.

\section{Lower bound on FIM of AoA and Numerical Results}\label{sec:results}
Since we are focused on analyzing the effects of ADC DR on AoA estimation, we evaluate the lower bound expression for the FIM in \eqref{eq:lowsnrfimbound1} for the AoA parameter, setting $\psi_i = \theta_{i}^{\mathrm{tar}}$ and $\psi_j = \theta_{j}^{\mathrm{tar}}$. The same process and assumptions can be applied to derive similar expressions for other parameters of interest.\useshortskip
\begin{align}
    &\left(\frac{\partial x_{n,\ell}}{\partial \theta_{i}^{\mathrm{tar}}} \right)^* 
    \frac{\partial x_{n,\ell}}{\partial \theta_{j}^{\mathrm{tar}}}= \nonumber \\
    &\left(\sum\limits_{m_1 \in \mathcal{C}_{0}}
    b_{m_1,0}^{(\ell)}\textbf{A}_{n,\ell,m_1}^{i,v}\textbf{f} 
    +\sum\limits_{k_1 = 1}^{K}\sum\limits_{m_2 \in \mathcal{C}_{k_1}}
    b_{m_2,k_1}^{(\ell)}\textbf{B}_{n,\ell,m_2}^{k_1,i,v}\textbf{f}^{\mathrm{u}}_{k_1}\right)^{*} \nonumber \\
    &\left(\sum\limits_{m_1^{'} \in \mathcal{C}_{0}}
    b_{m_1^{'},0}^{(\ell)}\textbf{A}_{n,\ell,m'_1}^{j,v}\textbf{f} 
    +\sum\limits_{k_2 = 1}^{K}\sum\limits_{m_2^{'} \in \mathcal{C}_{k_2}}
    b_{m_2^{'},k_2}^{(\ell)}\textbf{B}_{n,\ell,m'_2}^{k_2,j,v}\textbf{f}^{\mathrm{u}}_{k_2}\right),\label{eq:partialxnlthetamultiply}
\end{align}
where the terms are defined as follows.\useshortskip
\begin{align}
    &\textbf{A}_{n,\ell,m_1}^{i,v} = \Tilde{\gamma}_{i,\ell}^{\mathrm{tar}} c_{m_1}(2\tau_{i}^{\mathrm{tar}},v) \left(\dot{a}_{n,r}(\theta_{i}^{\mathrm{tar}})\textbf{a}_{t}^{\mathrm{T}}(\theta_{i}^{\mathrm{tar}})\right.\nonumber \\
    &\left.\;\;\;\;\;\;\;\;\;\;\;\;+ a_{n,r}(\theta_{i}^{\mathrm{tar}}) \dot{\textbf{a}}_{t}^{\mathrm{T}}(\theta_{i}^{\mathrm{tar}})\right),\\
    &\textbf{B}_{n,\ell,m_2}^{k_1,i,v} = \Tilde{\gamma}_{k_1,i,\ell}^{\mathrm{tar}} c_{m_2}(\phi_{k_1,i},v) \dot{a}_{n,r}(\theta_{i}^{\mathrm{tar}})\left(\textbf{a}_{k_1,t}^{\mathrm{u}}(\theta_{k_1,i}^{\mathrm{u}})\right)^{\mathrm{T}}.
\end{align}
Moreover, we use the following assumption
\begin{equation}\label{eq:assumption}
\mathbb{E}\left[b_{m_1,k_1}^{(\ell)*}b_{m_2,k_2}^{(\ell)}\right] 
=
\sigma^2_{k_1}
\delta_{m_1,m_2}
\delta_{k_1,k_2}.
\end{equation}
Under the assumption \eqref{eq:assumption} and by aggregating a large sample set, we simplify \eqref{eq:partialxnlthetamultiply} as follows.\useshortskip
\begin{align}
    &\left(\frac{\partial x_{n,\ell}}{\partial \theta_{i}^{\mathrm{tar}}} \right)^* 
    \frac{\partial x_{n,\ell}}{\partial \theta_{j}^{\mathrm{tar}}}= \sigma^2_{0}\sum\limits_{m_1 \in \mathcal{C}_{0}}
    \trace\left(\textbf{f}\textbf{f}^{\mathrm{H}}\left(\textbf{A}_{n,\ell,m_1}^{i,v}\right)^{\mathrm{H}}
    \textbf{A}_{n,\ell,m_1}^{j,v}\right) \nonumber \\
    &+\sum\nolimits_{\substack{k=1\\m_2 \in \mathcal{C}_{k}}}^{K}\sigma^{2}_{k}\trace
    \left(\textbf{f}^{\mathrm{u}}_{k}\left(\textbf{f}^{\mathrm{u}}_{k}\right)^{\mathrm{H}}
    \left(\textbf{B}_{n,\ell,m_2}^{k,i,v}\right)^{\mathrm{H}}\textbf{B}_{n,\ell,m_2}^{k,j,v}\right).
\end{align}
Substituting $\textbf{f}\textbf{f}^{\mathrm{H}} = \textbf{R}_{0}$ and $\textbf{f}^{\mathrm{u}}_{k}
\left(\textbf{f}^{\mathrm{u}}_{k}\right)^{\mathrm{H}} = \textbf{R}_{k}$, and accumulating over $NL$ samples, we arrive at \useshortskip
\begin{align}\label{eq:lowsnrfimboundthetaitarfinal}
    &\left(\frac{\partial \textbf{x}}{\partial \theta_{i}^{\mathrm{tar}}} \right)^H 
    \frac{\partial \textbf{x}}{\partial \theta_{j}^{\mathrm{tar}}} \nonumber \\
    &= \sum\nolimits_{n=1}^{N}\sum\nolimits_{\ell = 1}^{L}
    \left[\sigma^2_{0}\sum\limits_{m_1 \in \mathcal{C}_{0}}
    \trace \left(\textbf{R}_{0}\left(\textbf{A}_{n,\ell,m_1}^{i,v}\right)^{\mathrm{H}}
    \textbf{A}_{n,\ell,m_1}^{j,v}\right)\right. \nonumber \\
    &\left.+\sum\nolimits_{\substack{k=1\\m_2 \in \mathcal{C}_{k}}}^{K}\sigma^{2}_{k}\sum\limits_{}
    \trace\left(\textbf{R}_{k}\left(\textbf{B}_{n,\ell,m_2}^{k,i,v}\right)^{\mathrm{H}}\textbf{B}_{n,\ell,m_2}^{k,j,v}\right)\right].
\end{align}
Thus, the lower bound on the FIM for $\Theta$ is given as \useshortskip
\begin{align}
&\left[\textbf{F}_{\textbf{r}^{\mathrm{q}}}^{\mathrm{LS}}(\Theta)\right]_{ij}= \nonumber\\
&\frac{2(1-\eta)}{\sigma^2}\Re\left[
\sum\limits_{n=1}^{N}\sum\limits_{\ell = 1}^{L}\left(\sigma^2_{0}
\sum\limits_{m_1 \in \mathcal{C}_{0}}
\trace\left(\textbf{R}_{0}\left(\textbf{A}_{n,\ell,m_1}^{i,v}\right)^{\mathrm{H}}
\textbf{A}_{n,\ell,m_1}^{j,v}\right)\right.\right. \nonumber \\
&\left.\left.+\sum\nolimits_{\substack{k=1\\m_2 \in \mathcal{C}_{k}}}^{K}\sigma^2_{k}
\trace\left(\textbf{R}_{k}\left(\textbf{B}_{n,\ell,m_2}^{k,i,v}\right)^{\mathrm{H}}
\textbf{B}_{n,\ell,m_2}^{k,j,v}\right)\right)\right].\label{eq:fimthetalowsnr}
\end{align}
\subsection{Numerical Results}
We present simulation findings on the impact of ADC DR and quantization on HRF performance, focusing on the CRB for the AoA of targets. We use the FIM expression in \eqref{eq:fimthetalowsnr} to compute the upper bound on the CRB of the AoA. Simulations are conducted using a DFRC BS with one user ($K=1$) fixed at 100 meters (m) from the DFRC BS and one target ($P=1$), varying the target distance from 100 to 160 m. The system operates at 24 GHz, with 8 BS antennas and 4 user antennas. The system uses
14 OFDM symbols, with 36 subcarriers for DL and 24 for UL, each at a 15 kHz bandwidth. All parameters align with typical mmWave system values \cite{8207426}.
\begin{figure}[!t]
    \centering
        \includegraphics[width=6.2cm,height=6.2cm,keepaspectratio]{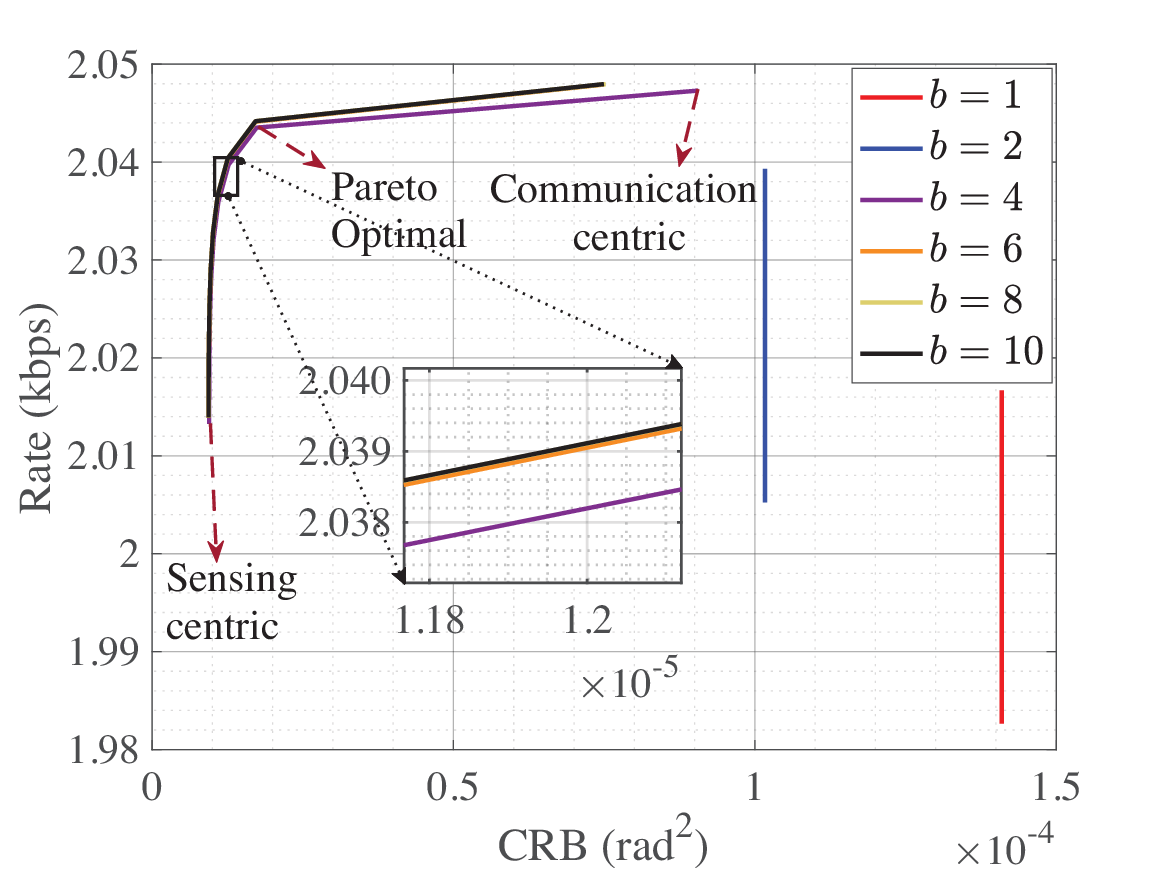}
        \caption{CRB vs UL rate for K = 1 and P = 1 varying the ADC resolution and fixing the user and target at a distance of 100 m from the DFRC BS.}\vspace{-0.1in}
        \label{fig:pld100}
\end{figure}
\begin{figure}[!t]
    \centering
        \includegraphics[width=6.2cm,height=6.2cm,keepaspectratio]{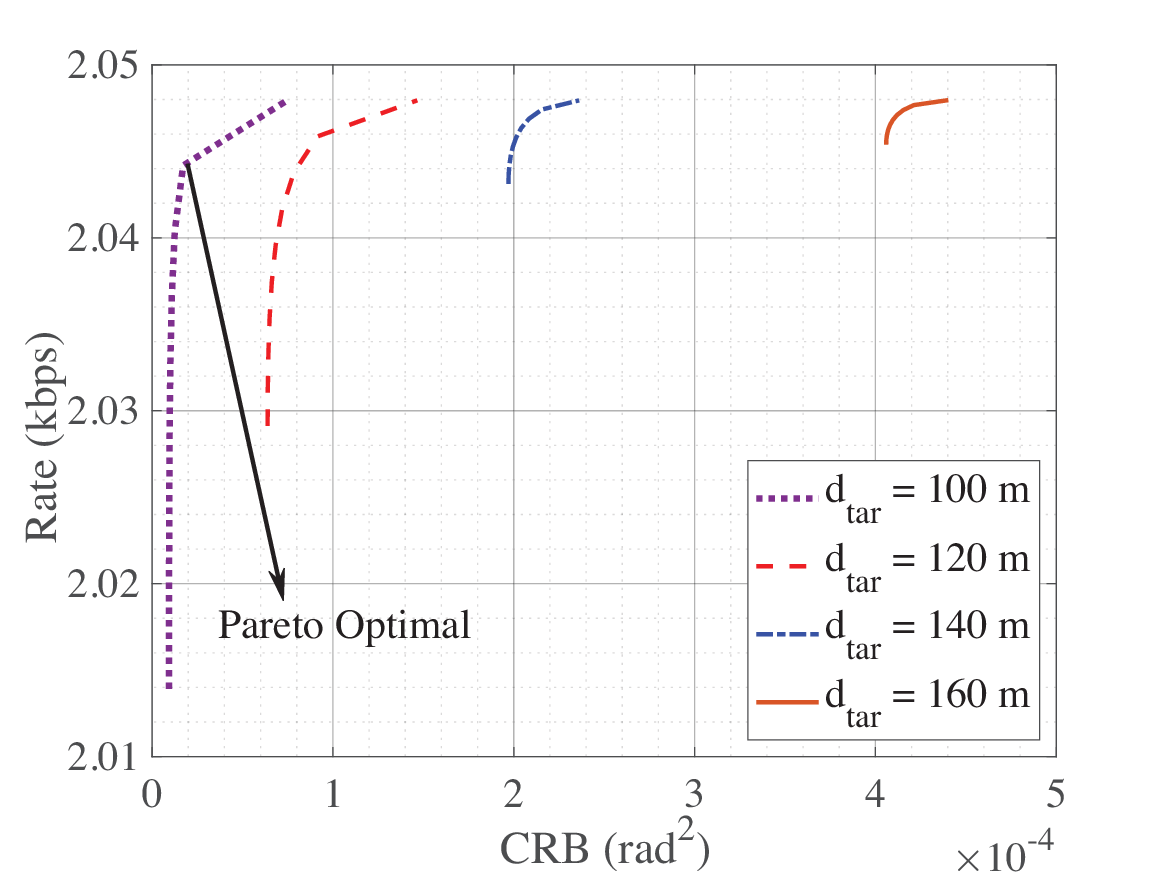}
        \caption{CRB vs. UL rate for K = 1 and P = 1, fixing the user and varying the distance of the target from the DFRC BS.}\vspace{-0.1in}
        \label{fig:pld}
\end{figure}
\begin{figure}[!t]
    \centering   
        \includegraphics[width=6.2cm,height=6.2cm,keepaspectratio]{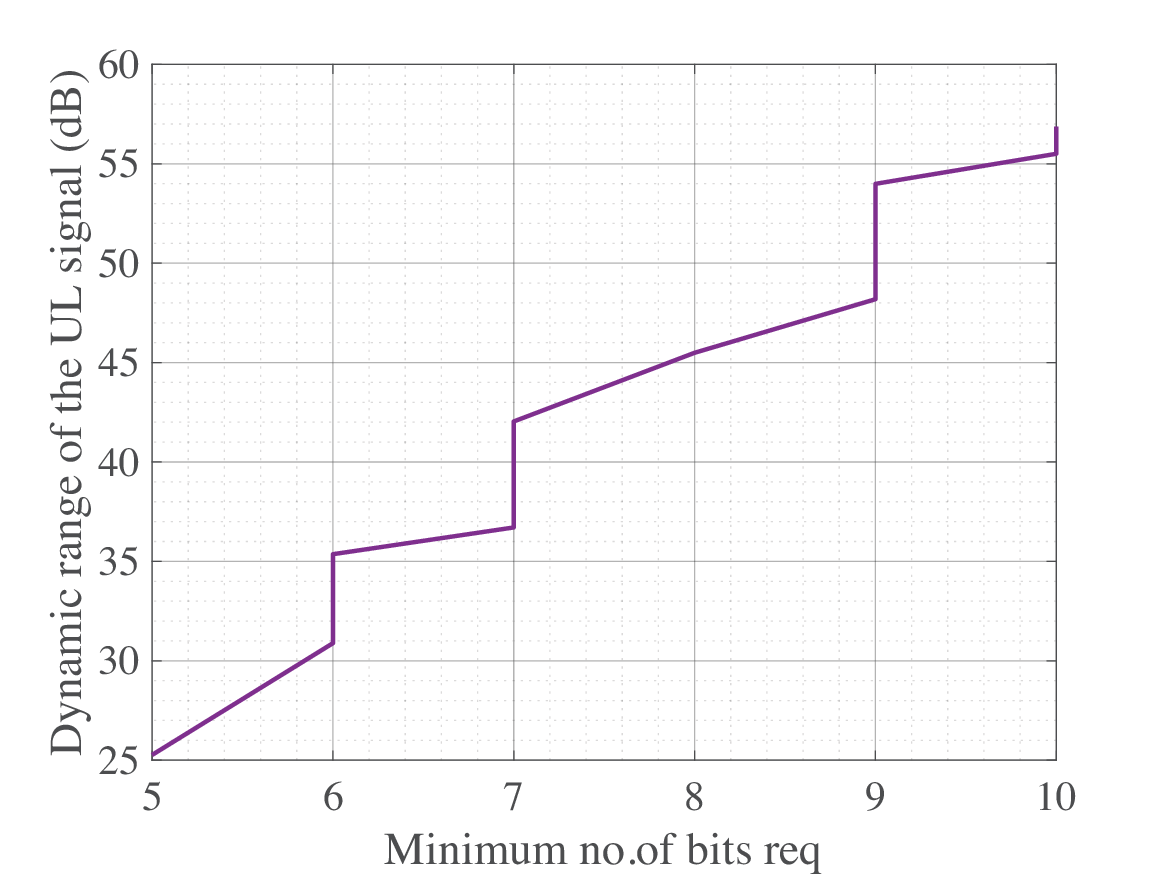}
        \caption{DR of the UL signal vs. the minimum required ADC resolution to differentiate the reflection and the direct path for K = 1 and P = 1.}
        \label{fig:dr}
\end{figure}
\subsubsection{CRB-Rate Trade-off with ADC DR} The simulation results presented in Fig.~\ref{fig:pld100} provide valuable insights into the CRB-rate trade-off in HRF systems, particularly with ADC DR. By fixing the target at $100$ m, we plot CRB-rate boundaries for various ADC resolutions. For each ADC setting, the sensing-centric point on the CRB-rate boundary is obtained by solving optimization problem $\mathbb{P}_0$ in \eqref{eq:P0_1} with $\mu = 0$, while the communication-centric point is determined by solving $\mathbb{P}_1$ in \eqref{eq:P11} with $\Gamma = 0$. The entire CRB-rate boundary is then constructed by varying $\mu$ in \eqref{eq:P05} across its range, as outlined in Section~\ref{sec:optmization}. These Pareto-optimal points represent the best achievable trade-offs between CRB and rate for each ADC resolution.

The ADC DR, as discussed in \cite{8617945}, scales with the ADC resolution, $b$, where higher resolution corresponds to greater DR. We plot the CRB-rate boundary across different values of $b$. The CRB remains unchanged for $b = 1$ and $b = 2$ bits, indicating that the ADC's DR is too low to separate the reflected signal from the noise floor. This is because the DR of the UL signal, defined as
\begin{equation}
    \text{DR}_{\text{sig}} = 10\log_{10}\left(\frac{P_{k}^{\mathrm{dp}}}{P_{k,j}^{\mathrm{ref}}}\right),\label{eq:dr_sig}
\end{equation}
where $P_{k}^{\mathrm{dp}}$ is the power of the direct path for user $k$ and $P_{k,j}^{\mathrm{ref}}$ represents the reflected signal power from user $k$ and target $j$, surpasses the ADC DR for these low resolutions. As a result, the weaker reflected signal is masked by the noise floor, preventing effective HRF processing at the DFRC BS. 

However, with an ADC resolution of $b = 4$ bits, the CRB starts to vary with rate, reflecting the improved ability of the ADC to differentiate the reflected signal from noise. This enables the DFRC BS to perform HRF successfully. The change in CRB with $\mu$ in $\mathbb{P}_0$ illustrates this; as $\mu$ varies, the power levels of the direct and reflected signals also change, which affects the signal's DR. In contrast, for lower resolutions ($b = 1$ and $b = 2$ bits), the CRB remains constant regardless of changes in $\mu$, underscoring the ADC's limitation in distinguishing the weaker signal even with adjusted DR values. For higher resolutions ($b \geq 4$ bits), the DR of the ADC increases sufficiently, allowing the BS to identify the weaker reflected signal and enabling an effective HRF.

Additionally, as $b$ increases, the CRB decreases, and the rate improves. However, beyond a certain resolution, further increases in $b$ yield only minimal improvements in the CRB and the rate, suggesting that additional bits become unnecessary.

\subsubsection{Impact of Target Distance on CRB-Rate Trade-off} The plot in Fig.~\ref{fig:pld} depicts the CRB-rate boundary as a function of the target's distance from the BS, assuming an ADC resolution of $b = 14$ bits. The results demonstrate that as the target distance from the BS increases, the CRB and the worst-case communication rate rise, keeping the user’s position constant. This trend primarily results from the increased attenuation of the reflected UL signal as the target moves further from the BS. Additionally, as the target distance from the BS grows, the distance between the user and the target increases, further amplifying the attenuation of the reflected UL signal utilized in HRF. Consequently, the CRB-rate boundary contracts as the power of the reflected signal diminishes. 

Furthermore, as the target distance extends from 100 m to 160 m, the Pareto-optimal points shift towards higher CRB values, underscoring an intensifying trade-off between CRB and rate with distance. At 100 m, the Pareto-optimal point occurs at the lowest CRB of approximately $0.1 \times 10^{-4} \, \text{rad}^2$ and a rate around 2.044 kbps. At 120 m, this balance shifts, yielding a CRB of $0.6 \times 10^{-4} \, \text{rad}^2$ and a rate of 2.045 kbps. For 140 m, the CRB increases to $1.5 \times 10^{-4} \, \text{rad}^2$, with a rate of around 2.047 kbps. Finally, at 160 m, the CRB reaches $3.5 \times 10^{-4} \, \text{rad}^2$, with a rate close to 2.047 kbps.

Conversely, the CRB decreases as the target moves closer to the BS. In this scenario, despite the reduced distance between the user and the target, the BS transmit power becomes the dominant factor, thus lowering the CRB and enhancing sensing performance. However, when the BS power is overly dominant, the HRF gains become marginal. This implies a trade-off between the echo power and the power of the reflected signal received at the BS. Similar observations are discussed in \cite{10417003}.

\subsubsection{Minimum Bits for ADC} Figure~\ref{fig:dr} shows the minimum ADC resolution required as a function of the DR of the UL signal received at the DFRC BS. This plot is created by positioning the target at random locations within a 200 m radius around the BS while the user is fixed at a distance of 100 m from the BS. At each target position, the DR of the UL signal received at the BS is calculated using \eqref{eq:dr_sig}, along with the minimum ADC bit resolution necessary to observe the HRF effect. The results indicate that as the DR of the UL signal received at the BS increases, the minimum required ADC resolution also rises, emphasizing the relationship between signal DR and the ADC resolution needed to perform HRF effectively.

\section{Conclusion and Future Work}
In this work, we have investigated the performance limits of HRF systems, considering the quantization effects of finite-resolution ADCs. Our analysis highlights the crucial role of ADC DR in determining CRB and communication rate trade-offs. We have derived the upper bound of quantized CRB for the estimation of AoA for HRF systems and established lower bounds on the FIM and UL communication rate. Two optimization problems were proposed to plot the CRB-rate boundaries, showing that increasing ADC resolution improves the system's ability to differentiate between direct and reflected signals, enhancing both sensing accuracy and communication quality. Future work focuses on developing beamforming techniques to mitigate ADC DR limitations, with the aim of reducing the required ADC resolution and optimizing overall system performance. This approach paves the way for more advanced ISAC systems that balance communication and sensing under hardware constraints.
\vspace{-0.1cm}§
\section*{Acknowledgment}

This work is supported by Tamkeen under the Research Institute NYUAD grant CG017.

\bibliographystyle{IEEEtran}
\bibliography{ref}
\end{document}